\documentclass{ws-mpla}

\newcommand{\asmz}{\alpha_s(M_Z)}

\begin{document}

\markboth{A M Cooper-sarkar}
{Determination of charm quark mass and $\asmz$ from HERA data}

\catchline{}{}{}{}{}

\title{DETERMINATION OF CHARM QUARK MASS AND $\asmz$ FROM HERA DATA}

\author{\footnotesize AMANDA COOPER-SARKAR}

\address{Dept. of Physics, University of Oxford,Keble Rd, OXFORD, OX1 3RH, UK\\
email:a.cooper-sarkar1@physics.ox.ac.uk}

\maketitle

\pub{Received (24 May 2013)}{Revised ()}

\begin{abstract}
Charm production data from HERA may be used to determine the charm quark mass 
and jet production data from HERA may be used to detrmine $\asmz$. 
Recent results are summarised.
\end{abstract}

\section{Introduction}

HERA was an electron(positron)-proton collider located at DESY, Hamburg. 
It ran in two phases HERA-I from 1992-2001 and HERA-II 2003-2007. 
Two similar experiments, H1 and ZEUS, took data. In HERA-1 running each 
experiment collected  $\sim 100$pb$^{-1}$ of $e^+p$ data 
and $\sim15$pb$^{-1}$ of $e^-p$ data with electron beam energy $27.5$GeV and 
proton beam energies $820,920~$GeV. In HERA-II running each experiment took 
$\sim 140$pb$^{-1}$ of $e^+p$ data 
and $\sim180$pb$^{-1}$ of $e^-p$ data with the same electron beam energy and 
proton beam energies $920~$GeV. 
	
Deep inelastic lepton-hadron 
scattering data has been used both to investigate the theory of 
the strong interaction and to determine the momentum distributions of the 
partons within the nucleon. The data from the HERA collider now dominate 
the world data on deep inelastic scattering since they cover an unprecedented 
kinematic range: in $Q^2$, the (negative of the) 
invariant mass squared of the virtual exchanged boson, 
$0.045 < Q^2 < 3\times 10^{-5}$; in Bjorken $x$, $6\times10^{-7} < x < 0.65$.
Futhermore, because the HERA experiments investigated
 $e^+p$ and $e^-p$, charge current(CC) and neutral current (NC) scattering, 
information can be gained on flavour separated up- and down-type quarks and 
antiquarks and on the gluon- from its role in the scaling violations of 
perturbative quantum-chromo-dynamics. The formalism of how the deep inelastic 
cross sections relate to the parton distributions though the QCD-improved 
parton model, and how parton didtribution functions are then extracted 
from the data, is well documented (see for example ref.~\cite{jphysgrev}) and 
only a brief description of the HERAPDF formalism will be given here. 
This contribution concentrates on results of relevance 
for the fundamental parameters of QCD: $\alpha_s(M_Z)$ and the charm qurak 
mass $m_c$.               
  
\section{HERAPDF Formalism}
Perturbative QCD predicts the $Q^2$ evolution of the parton distributions, but
not the $x$ dependence.
Parton distributions are extracted by 
performing a direct numerical integration of the DGLAP evolution 
equations at NLO.  
A parametrised analytic shape for the parton distributions 
 is assumed to be valid  at some starting value of $Q^2 = Q^2_0$. 
For the HERAPDF the value $Q^2_0 = 1.9~$GeV$^2$ is chosen such 
that the starting scale is below the charm mass threshold, $Q_0^2 < m_c^2$. 
Then the DGLAP equations
are used to evolve the parton distributions up to a different $Q^2$ value, 
where they  are convoluted with NLO 
coefficient functions to make predictions 
for the structure functions.  The heavy quark 
coefficient functions are usually calculated in the general-mass
variable-flavour-number scheme of Thorne~\cite{Thorne:1997ga}. 
However, for the study of the charm mass 
various other schemes have been considered, see Section~\ref{sec:charm} 
The heavy quark masses  for the central HERAPDF fits 
were chosen to be $m_c=1.4~$GeV 
and $m_b=4.75~$GeV and the strong coupling constant was fixed to 
$\alpha_s(M_Z) =  0.1176$. (The values of $m_c$ and $\asmz$ are varied when 
studies of these parameters are performed.) 
The predictions are then fitted to the 
combined HERA data sets for NC and CC $e^+p$ and $e^-p$ scattering. 
A minimum $Q^2$ cut of $Q^2_{min} = 3.5$~GeV$^2$ was imposed 
to remain in the kinematic region where
perturbative QCD should be applicable.

 The fit parameters are those 
necessary to specify the input analytic shape.
The valence quark 
distributions $xu_v$, $xd_v$, and the $u$-type and $d$-type anti-quark 
distributions
$x\bar{U}$, $x\bar{D}$ ($x\bar{U} = x\bar{u}$, 
$x\bar{D} = x\bar{d} +x\bar{s}$) are parametrised at the input scale 
$Q^2_0=1.9$GeV$^2$ by the generic form 
\begin{equation}
 xf(x) = A x^{B} (1-x)^{C} (1 + D x + E x^2).
\label{eqn:pdf}
\end{equation}
The parametrisation for the gluon distribution $xg$ can be extended, 
to include a 
term $-A_{g'} x^{B_{g'}} (1-x)^{C_{g'}}$, such that the NLO gluon may become 
negative at low $x$ and low $Q^2$ (however it does not do so in the kinematic 
region of the HERA data). 
The normalisation parameters, $A_g, A_{u_v}, A_{d_v}$, are constrained 
by the quark number sum-rules and momentum sum-rule. Constraints are 
imposed to ensure that $\bar{u}=\bar{d}$ at low-x, and to specify 
the strangeness fraction in the sea.
The full details for all HERAPDFs are given in ref.~\cite{jphysgrev}

The experimental uncertainties on the HERAPDF are determined using 
 the conventional 
$\chi^2$ tolerance, $\Delta\chi^2=1$, for $68\%$C.L. However model 
uncertainties and parametrisation uncertainties are also considered.
 The choice of the heavy 
quark masses is varied,
the choice of $Q^2_{min}$ is varied and 
the strangeness fraction is varied.
Parametrisation variations for which
 the $E$ and the $D$ parameters for all the 
PDFs are freed are considered and variation of the 
starting scale $Q^2_0$ in the range, $1.5 < Q^2_0 < 2.5$GeV$^2$, is also 
considered. These model and parametrisation uncertainties are an 
integral part of the HERAPDF uncertainties.

\section{Results}
From 2008, the H1 and ZEUS experiments 
began to combine their data in order to provide the most complete and 
accurate set of deep-inelastic data as the legacy of HERA.
 Data on 
inclusive cross-sections have been combined for the HERA-I phase of 
running~\cite{h1zeus:2009wt} 
and a preliminary combination has been made also using the HERA-II 
data~\cite{highq2}.
HERA-jet data have been added to this combination in order to give more 
information on $\alpha_s(M_Z)$ and the gluon distribution~\cite{herapdf16}. 
Very recently a combination of data on  $F_2^{c\bar{c}}$ has been used 
to give information on the charm quark mass~\cite{charmcomb}.
 
\begin{figure}[tbp]
\vspace{-0.5cm} 
\centerline{
\epsfig{figure=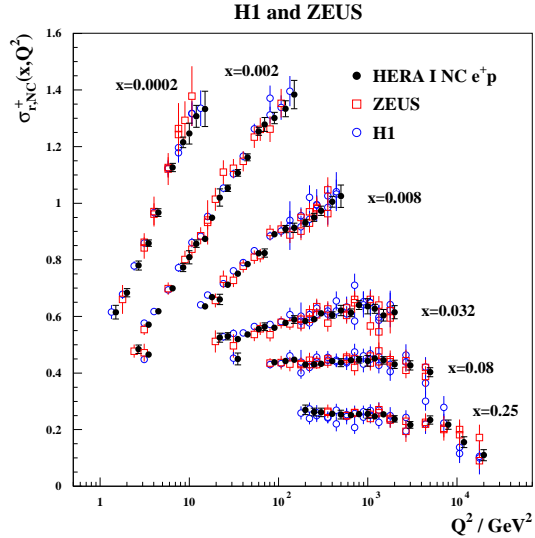,height=0.4\textheight}}
\caption {HERA combined NC $e^+p$ reduced 
cross section as a function of 
$Q^2$ for six $x$-bins compared to the separate 
H1 and ZEUS data input to the averaging procedure.
The individual measurements are displaced horizontally for a better 
visibility.
}
\label{fig:quality}
\end{figure}

 The combination of the H1 and ZEUS data sets 
takes into account the full correlated systematic uncertainties of the 
individual experiments such that the total uncertainty of the combined measurement is typically smaller than 
$2\%$, for $3 < Q^2 < 500$~GeV$^2$, and reaches $1\%$, for $20 <  Q^2 < 
100$~GeV$^2$. 
In Fig~\ref{fig:quality} averaged data are compared to the 
input H1 and ZEUS data, illustrating the improvement in precision. Because 
of the reduction in size of the systematic error this improvement is far 
better than would be expected simply from the rough doubling of statistics 
which combining the two experiments represents. 

A combination has also been made of data on $F_2^{c\bar c}$~\cite{charmcomb} from 
various different methods of tagging charm: using the $D^*$, using 
the vertex detectors to see the displaced decay vertex, using direct 
$D_0, D^+$ production identified using the vertex detectors, and indentifying 
semi-leptonic charm decays via muons, also using the vertex detectors.

\begin{figure}[tbp]
\vspace{-0.5cm} 
\centerline{
\epsfig{figure=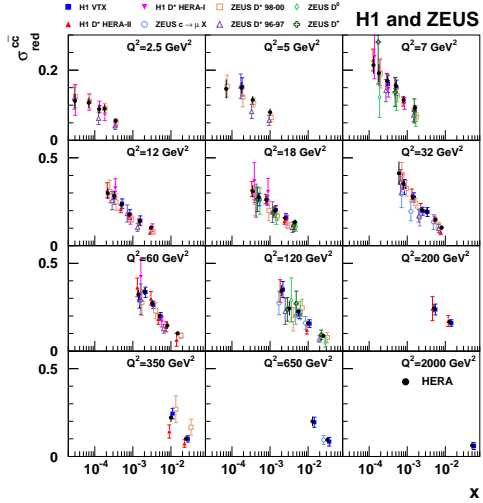,height=0.35\textheight}}
\caption {The HERA combined measurement of $F_2^{c\bar c}$ compared to the 
data sets of H1 and ZEUS used for the combination. these data sets are 
slightly displaced in $x$ for visibility.
}
\label{fig:f2ccomb}
\end{figure}

\begin{figure}[tbp]
\vspace{-0.5cm} 
\centerline{
\epsfig{figure=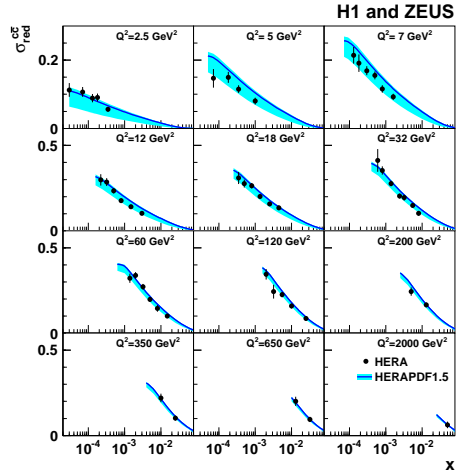,height=0.35\textheight}}
\caption {The HERA combined measurement of $F_2^{c\bar c}$ compared to the 
predictions of HERAPDF1.0
}
\label{fig:f2cfit}
\end{figure}

The results of the $F_2^{c\bar c}$ combination compared to the 
separate measurements which 
went into it are shown in Fig~\ref{fig:f2ccomb}.  
The  $F_2^{c\bar c}$ combination is shown compared to the predictions of 
HERAPDF1.0 in Fig.~\ref{fig:f2cfit}. 

\subsection{Charm quark mass}
\label{sec:charm}
\begin{figure}[tbp]
\vspace{-0.5cm} 
\begin{tabular}{cc}
\includegraphics[width=0.42\textwidth]{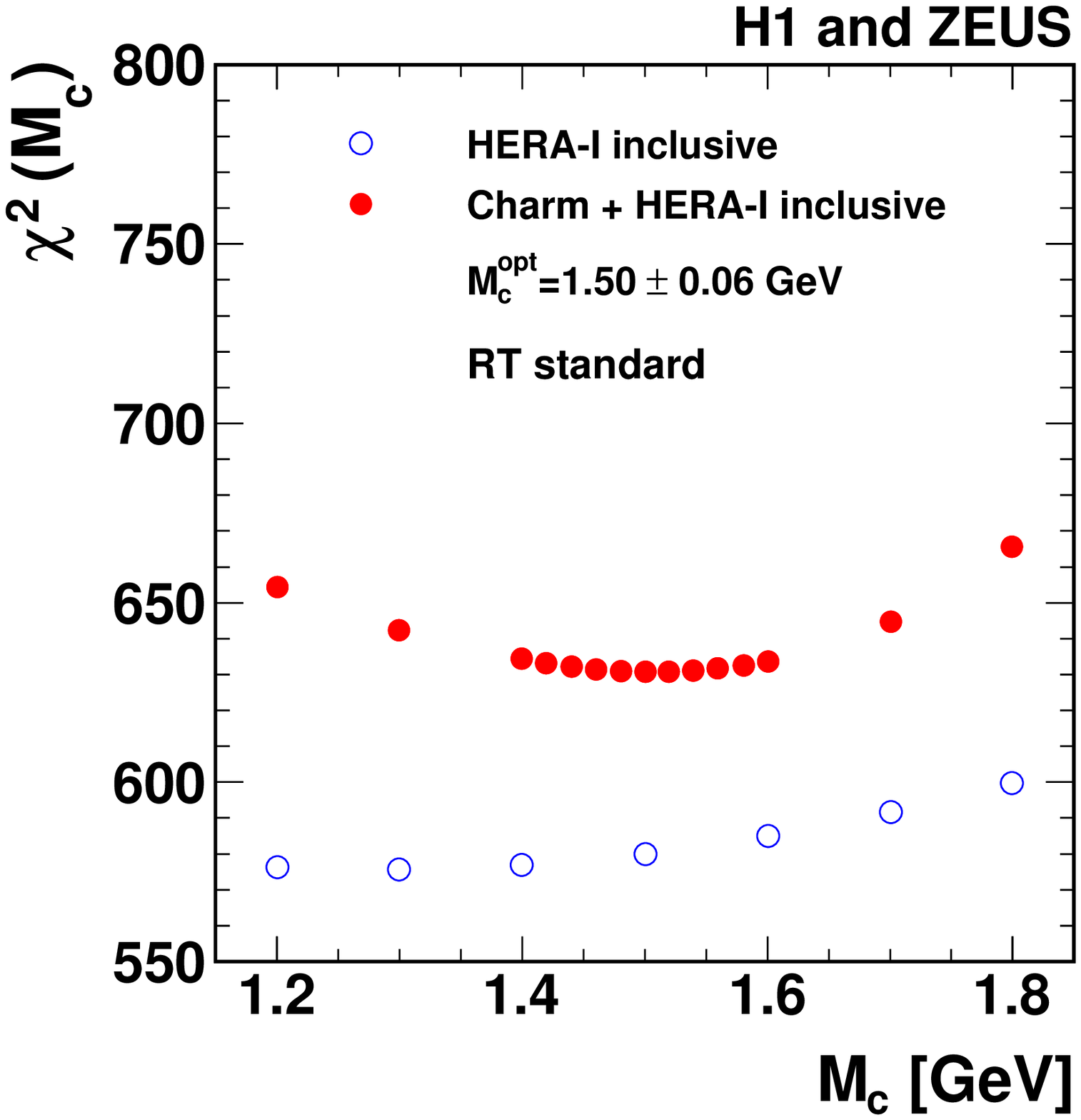} &
\includegraphics[width=0.42\textwidth]{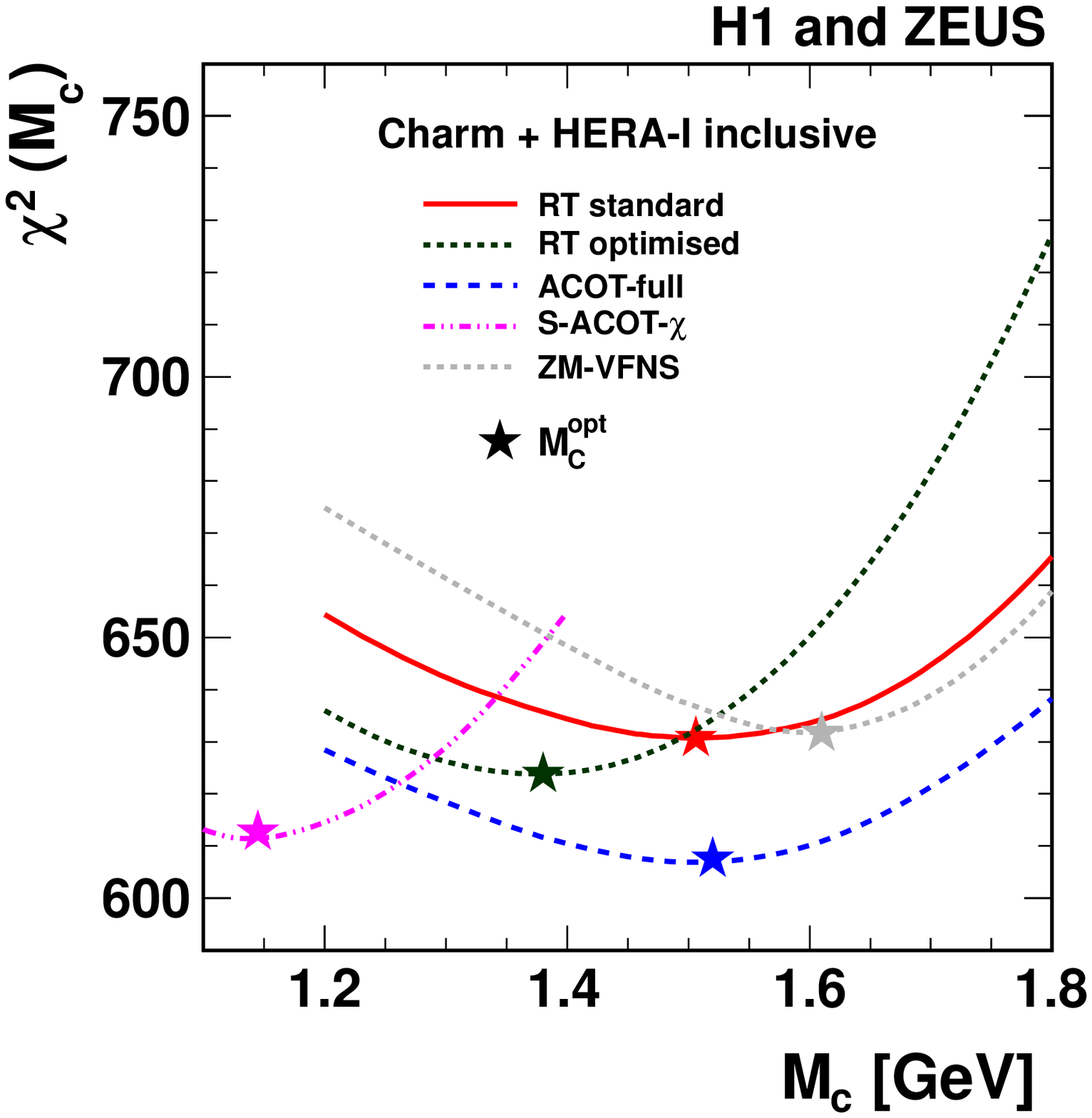} 
\end{tabular}
\begin{tabular}{cc}
\includegraphics[width=0.42\textwidth]{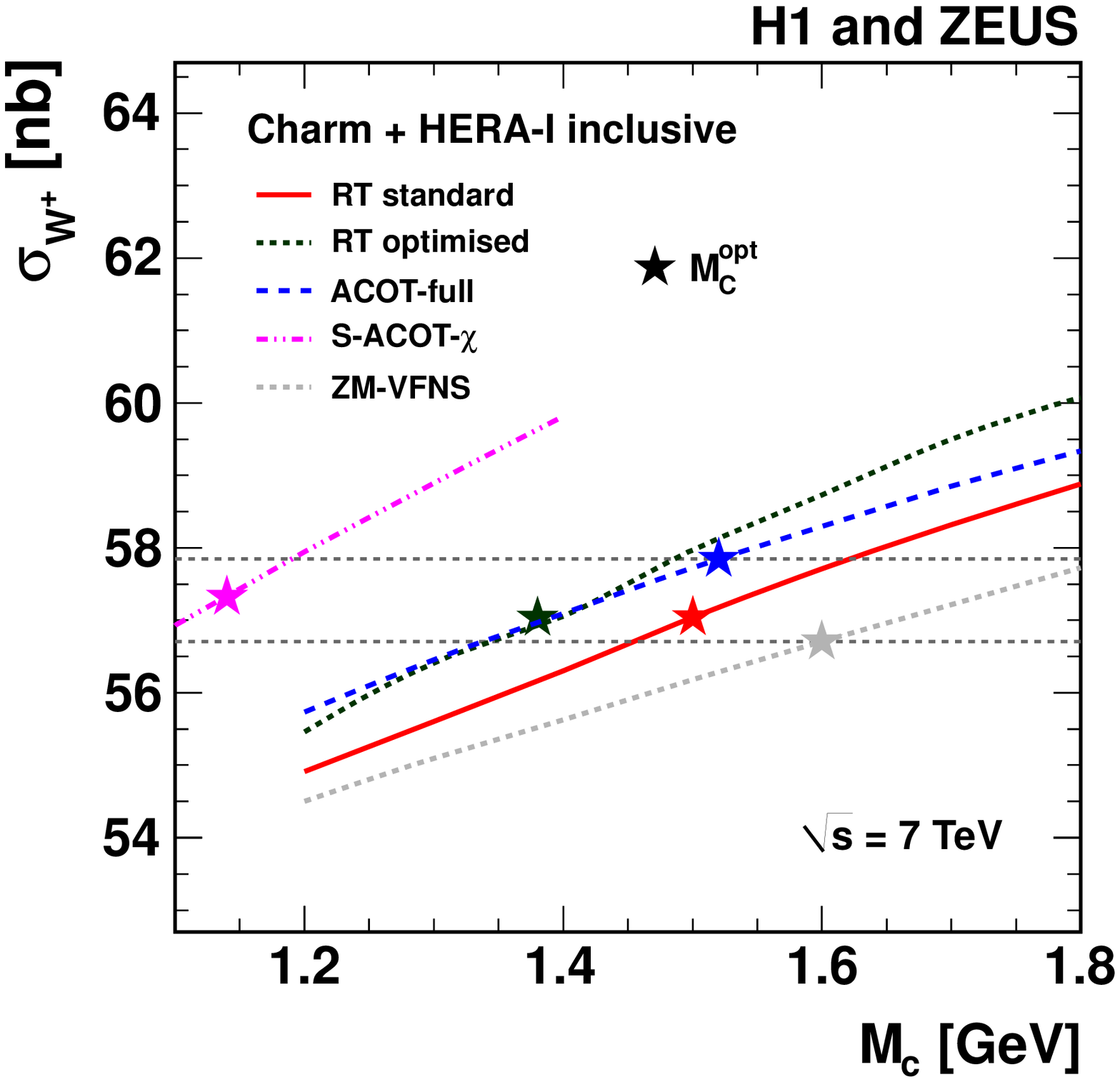} &
\includegraphics[width=0.42\textwidth]{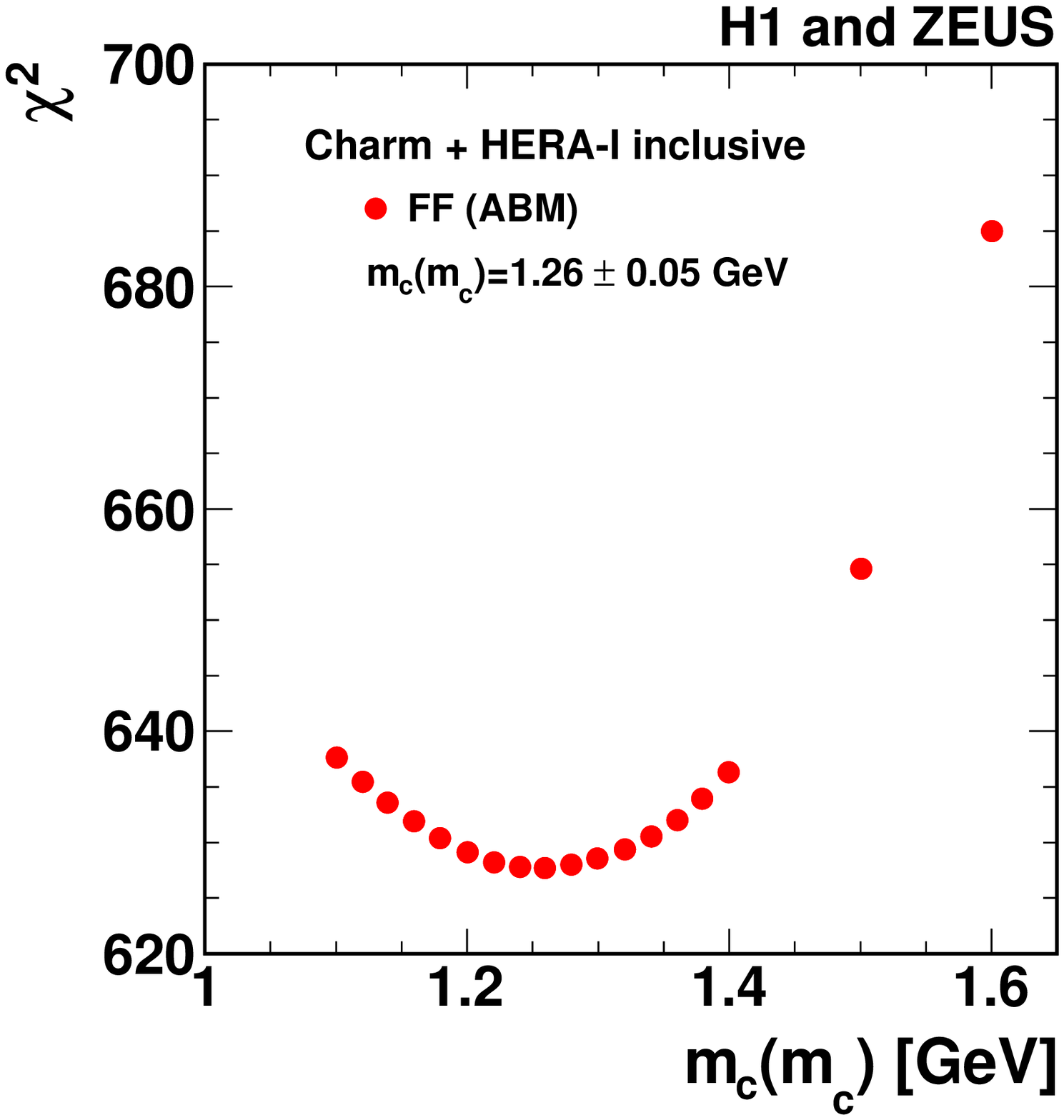}
\end{tabular}
\caption {The $\chi^2$ of the HERAPDF fit as a 
function of the charm mass parameter $m_c^{model}$. Top left; using the 
RT-standard heavy-quark-mass scheme, when only inclusive DIS data are included 
in the fit and 
when the data for $F_2^{c\bar{c}}$ are also included in the fit. Top right;
 using various heavy-quark-mass schemes, when the data for $F_2^{c\bar{c}}$
 are also included in the fit. Bottom left; predictions for the $W^+$ 
cross-sections at the LHC, as a 
function of the charm mass parameter $m_c^{model}$, for various 
heavy-quark-mass schemes. Bottom right: Using the FFN scheme in terms of 
the running mass $m_c(m_c)$.
}
\label{fig:charmpred}
\end{figure}

The uncertainty bands on these predictions 
are dominated by the variation of the  charm quark mass, 
$ 1.35 < m_c < 1.65$GeV, used for the model uncertainty. The combined $F_2^{c\bar{c}}$ data can help to reduce 
this uncertainty. 
Fig.~\ref{fig:charmpred} (top left) compares the $\chi^2$, 
as a function of the charm mass, for a fit which includes charm data 
to that for the HERAPDF1.0 fit which does not use these data, when using the 
Thorne-Roberts (RT) variable-flavour-number (VFN) scheme. The charm data have 
a clear preference for a particular charm quark mass. 
However, the RT heavy flavour VFN scheme is not unique,
specific choices are made for threshold behaviour. In Fig.~\ref{fig:charmpred} 
(top right) the $\chi^2$ profiles for the standard and the 
optimized versions (optimized for smooth 
threshold behaviour) of this scheme are compared. 
The same figure also compares the alternative ACOT VFN 
schemes and the Zero-Mass VFN scheme. Each of these schemes 
favours a different value for the charm quark mass, 
and the fit to the data is good for all the heavy-quark-mass 
schemes considered. 

Each of the VFN schemes has been used to predict $W$ 
and $Z$ production for the LHC and their predictions for $W^+$ are shown in 
Fig.~\ref{fig:charmpred} (bottom left) as a function of the charm quark mass. 
If a particular
value of the charm mass is chosen then the spread of predictions is as large as
$\sim7\%$.
However this spread is considerably reduced $\sim 1\%$ if each heavy quark 
scheme is used at its own favoured
value of the charm quark-mass.  Furher details of this study are 
given in ref.~\cite{charmcomb}.

It is important to note that the schemes so far considered are General Mass 
VFN Schemes and that the charm quark mass used in such 
schemes is the pole-mass. However since the since the relationship between the 
pole-mass and the running-mass (or MSbar mass) does not converge it is 
better to use the running-mass. This has been done within the Fixed Flavour 
Number scheme~\cite{abm} and the result is 
$
m_c(m_c)= 1.26\pm 0.05
$
as shown in Fig.~\ref{fig:charmpred} (bottom right).

\subsection{$\alpha_s(M_Z)$}

The value of $\alpha_s(M_Z)$ is strongly correlated to the gluon PDF shape 
in fits to
 inclusive DIS data, because the gluon PDF is determined indirectly 
from the scaling violations. 
Jet production cross-sections depend directly on the gluon PDF. 
This extra information reduces the strong correlation between the gluon and 
$\asmz$ and allows competitive measurements of $\asmz$. Such studies have 
been made using jet data alone and by making a simultaneous fit of PDFs 
and $\alpha_s(M_Z)$.

\begin{figure}[tbp]
\vspace{-1.0cm} 
\begin{center}
\includegraphics[width=0.72\textwidth]{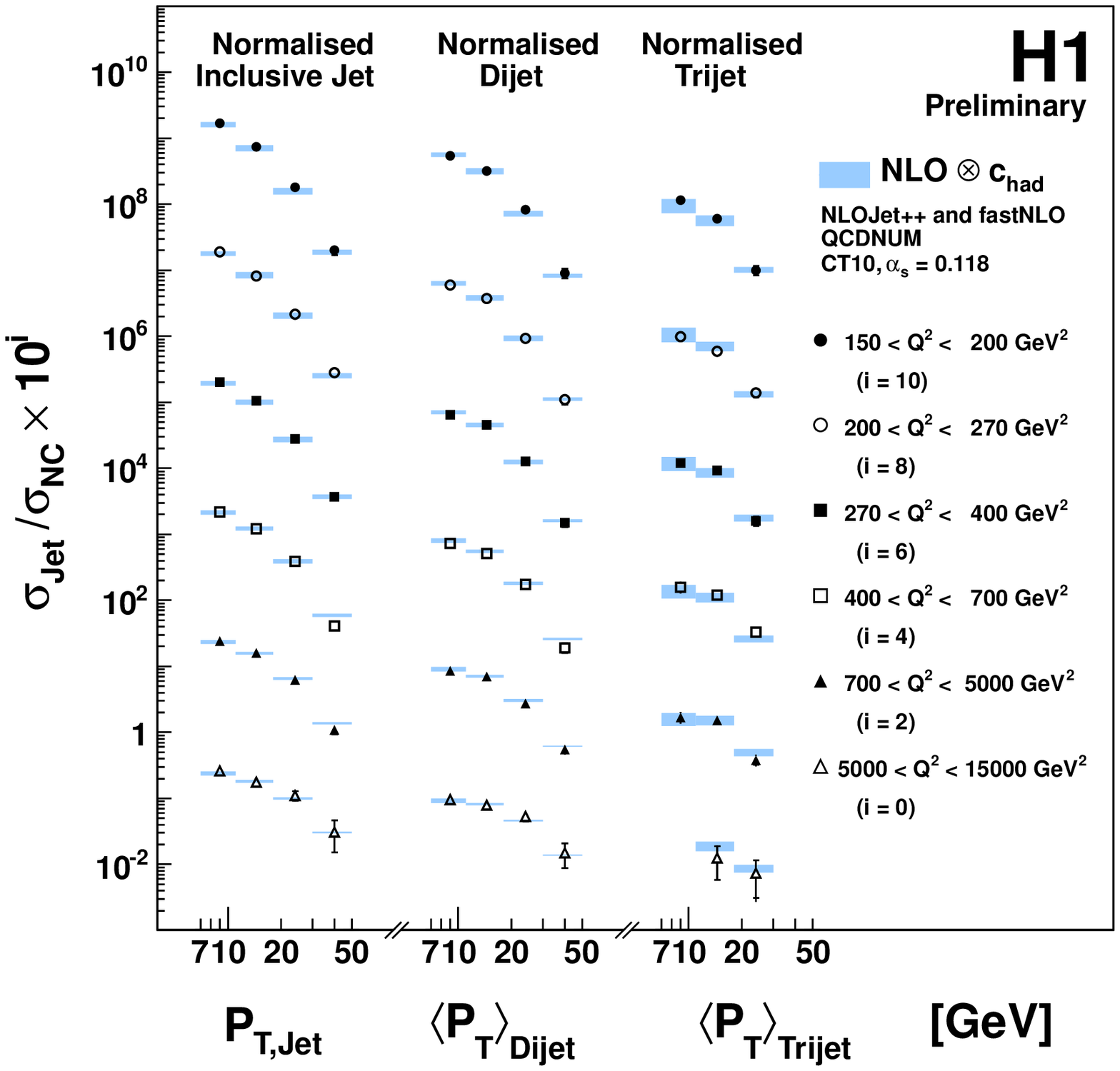}
\caption{Normalised inclusive jet,dijet and tri-jet cross sections measured by 
the H1 Collaboration as a function of the transverse momentum in different 
$Q^2$ bins compared to the NLO QCD predictions.}
\label{fig:h1jets}
\end{center}
\end{figure}

Two recent studies using jet data alone are reported here. 
The first uses H1 normalised 
inclusive jet, di-jet and tri-jet cross-sections~\cite{H1-prelim-12-031} 
which are well described by NLO QCD predictions from NLOJet++~\cite{NLOjet+}
 using CT10 PDFs~\cite{ct10} with $\asmz=0.118$, 
see Fig.~\ref{fig:h1jets}. 
A regularized unfolding is performed for all
bins and multijet categories simultaneously to correct
for detector effects and this yields a complete correlation matrix, which is 
used when these data are included into QCD fits which fix the PDF used 
(CT10 PDF) while varying $\asmz$.
The covariance matrix was used to normalize the different
jet rates to the inclusive neutral-current DIS cross section. 
This normalization reduces both
the experimental and the theoretical uncertainties.
Values of $\asmz$ are extracted separately from inclusive jets, di jets and 
tri-jets samples yielding:

$\alpha_S (M_Z)$ = 0.1197 $\pm$ 0.0008 (exp.) $\pm$ 0.0014 (PDF) $\pm$ 0.0011 (hadr.) $\pm$ 0.0053 (th.) for inclusive jets

$\alpha_S (M_Z)$ = 0.1142 $\pm$ 0.0010 (exp.) $\pm$ 0.0016 (PDF) $\pm$ 0.0009 (hadr.) $\pm$ 0.0048 (th.) for dijets and

$\alpha_S (M_Z)$ = 0.1185 $\pm$ 0.0018 (exp.) $\pm$ 0.0013 (PDF) $\pm$ 0.0016 (hadr.) $\pm$ 0.0042 (th.) for tri-jets.

There is some tension between the di-jet and inclusive jet measurements, 
possibly due to missing higher order calculations, so that to make a combined 
$\asmz$ extraction the cross-section predictions at NLO and 
LO were required to agree within 30\%. This was fulfilled for 42 out of 65 bins. The combined fit then gives

$\alpha_S (M_Z)$ = 0.1163 $\pm$ 0.0011 (exp.) $\pm$ 0.0014 (PDF) $\pm$ 0.0008 (hadr.) $\pm$ 0.0039 (th.),

 where the largest uncertainty is the theoretical 
uncertainty from scale variation.

\begin{figure}[tbp]
\vspace{-1.0cm} 
\begin{center}
\begin{tabular}{cc}
\includegraphics[width=0.48\textwidth]{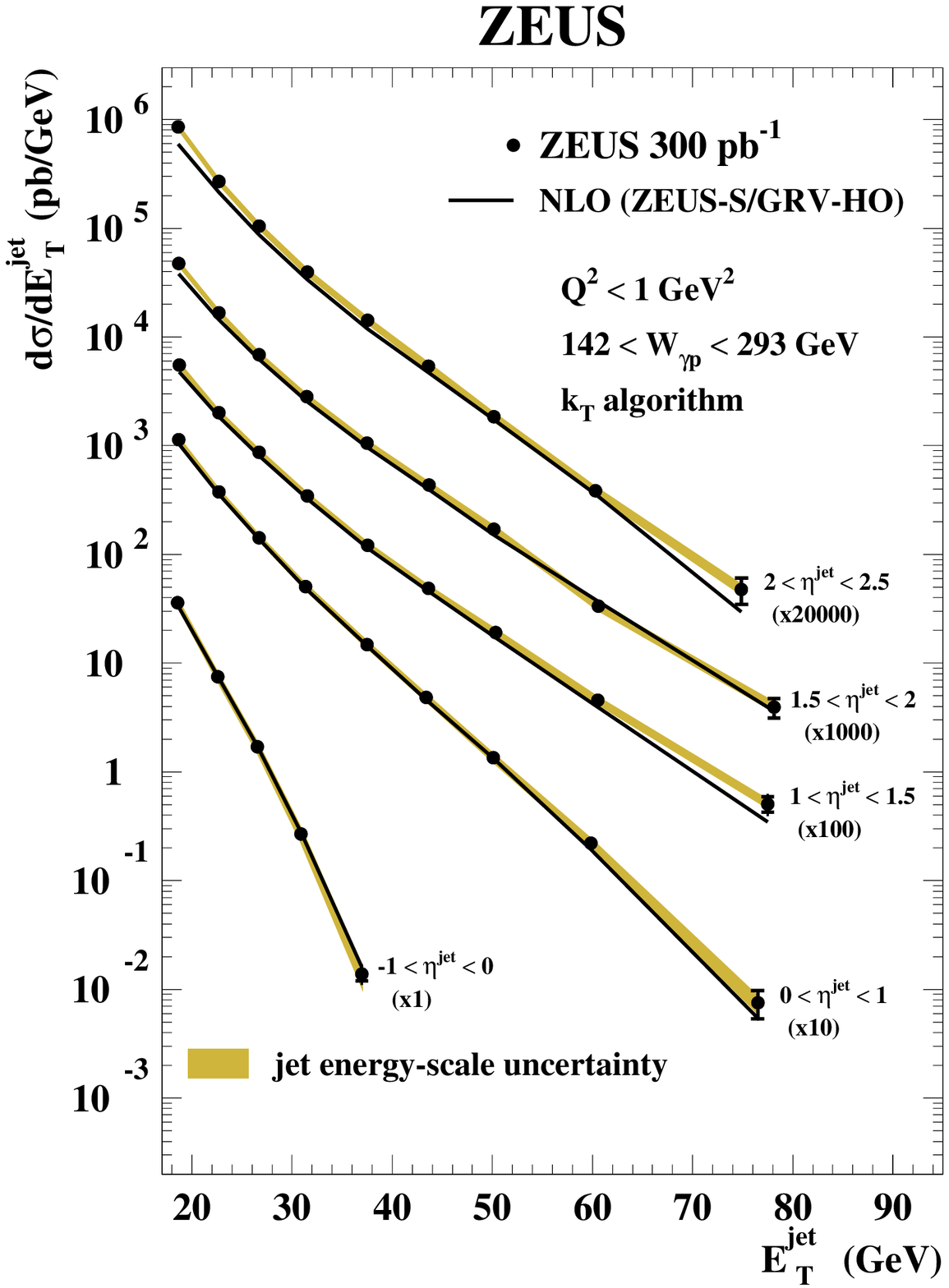} &
\includegraphics[width=0.48\textwidth]{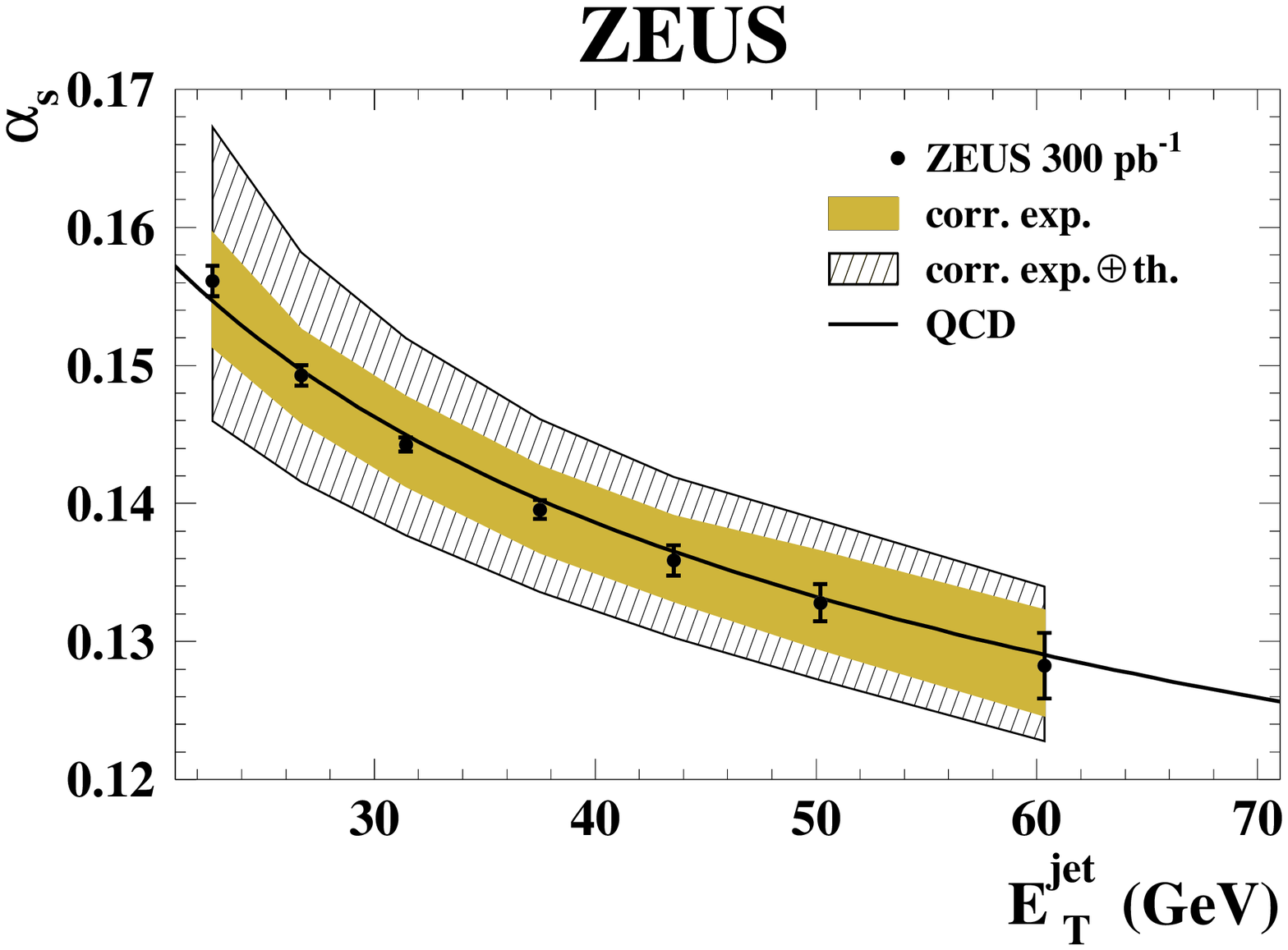}
\end{tabular}
\caption{Left: Inclusive jet cross section in photoproduction measured by the 
ZEUS Collaboration as a function of
  the transverse jet energy in different rapidity regions and compared to the 
NLO prediction. Right: the extracted values of $\alpha_s$ measured for 
different ranges of $E_T^{jet}$. }
\label{fig:zeusjets}
\end{center}
\end{figure}

The second study uses ZEUS photoproduction data, 
where a quasi-real photon is emitted from the incoming lepton.
In direct photoproduction, the photon directly takes part in the interaction, 
whereas
the photon acts as a source of partons in events with resolved photoproduction.
The ZEUS Collaboration published new double-differential jet cross sections in 
photoproduction events
with a center-of-mass energy of the photon-proton system between 142 and 
293~GeV \cite{Abramowicz:2012jz}.
Compared to DIS measurements, this analysis has a relatively high reach of 
transverse jet energies, $E_T$, up to 80~GeV.
Overall, a reasonable agreement between data and the NLO predictions 
(using ZEUS-S PDFs~\cite{zeuss}
 for the proton PDF, GRV-HO for the photon PDF and the 
programme of Klasen, Kleinwort and Kramer~\cite{EpJC1(1998)1}) is 
observed as shown in Fig.~\ref{fig:zeusjets}.
The ZEUS Collaboration has used these jet cross sections to measure $\asmz$.
The range in $d \sigma / d E_T$ was restricted to 21 $< E_T <$ 71~GeV in order 
to reduce
potential non-perturbative effects and the impact of the proton PDF 
uncertainty.
An $\alpha_S (M_Z)$ value of 

$\alpha_S (M_Z) = 0.1206\ ^{+0.0023}_{-0.0022}\ {\rm (exp.)}\ ^{+0.0042}_{-0.0035}\ {\rm (th.)}$

was obtained, where scale uncertainties again represent the dominant 
uncertainty. The $\asmz$ extraction is also performed for various different 
ranges of $E_T^{jet}$ and this demonstrates the running of $\alpha_s$ within 
a single experiment, as seen in Fig.~\ref{fig:zeusjets}.

\begin{figure}[tbp]
\vspace{-1.0cm} 
\begin{center}
\begin{tabular}{cc}
\includegraphics[width=0.48\textwidth]{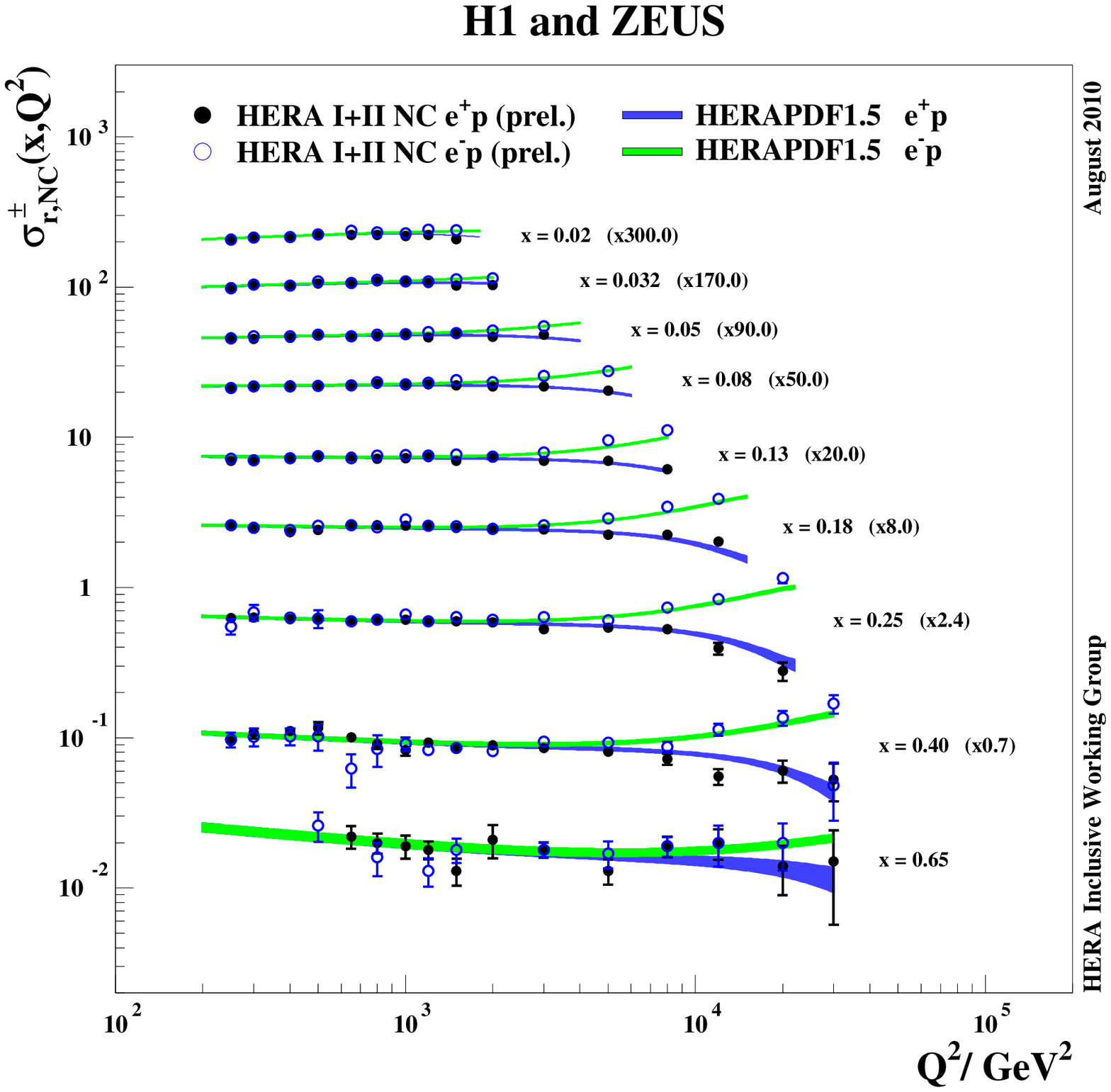} &
\includegraphics[width=0.48\textwidth]{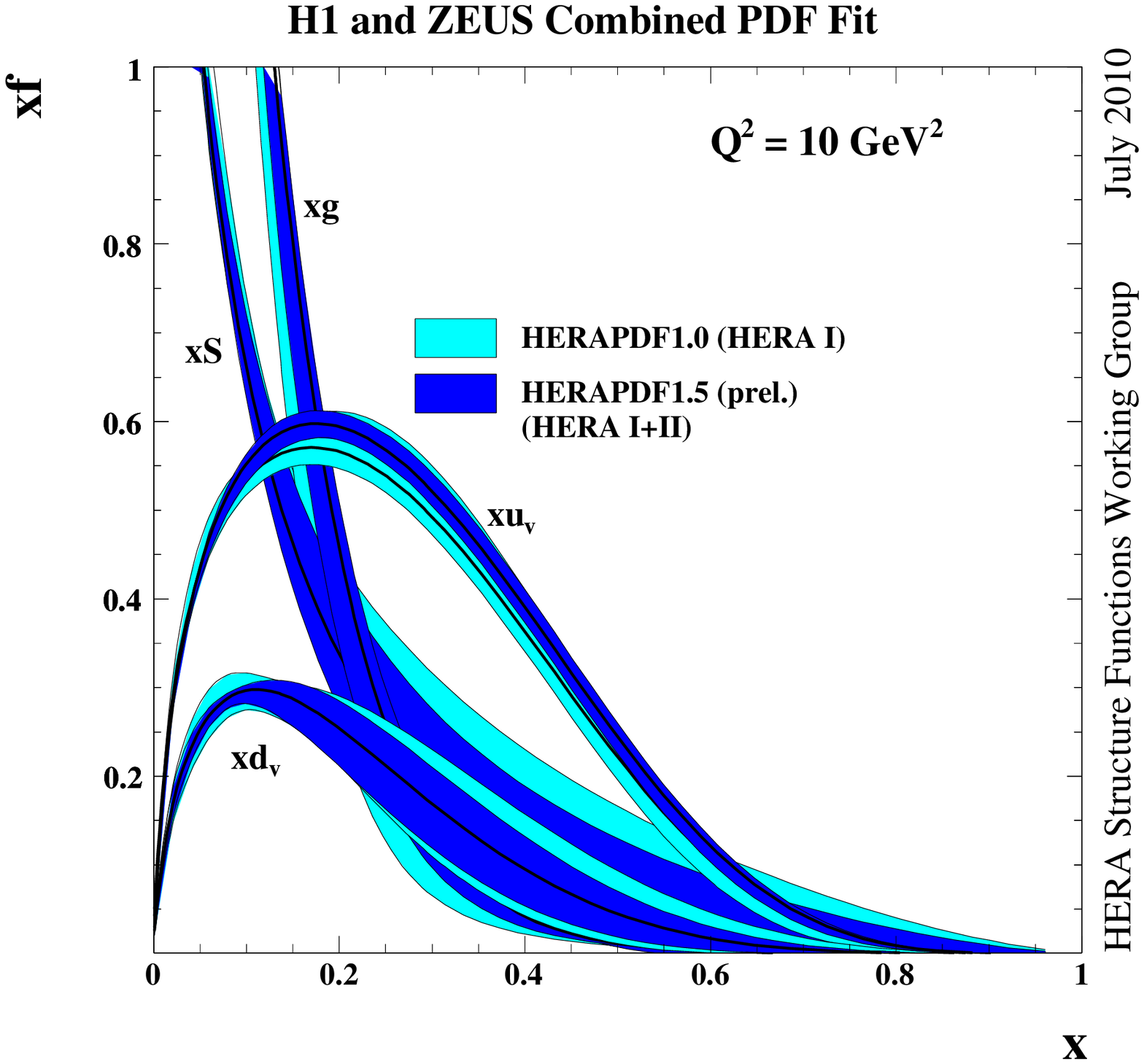}
\end{tabular}
\caption{Left: HERA combined data points for the NC $e^{\pm}p$ cross-sections 
as a function of $Q^2$ in bins of $x$, for data from the HERA-I and II run 
periods. The HERAPDF1.5 fit to these data is also shown on the plot.
Right: Parton distribution functions from HERAPDF1.0 and HERAPDF1.5; $xu_v$, 
$xd_v$,$xS=2x(\bar{U}+\bar{D})$ and $xg$ at $Q^2=10~$GeV$^2$.}
\label{fig:herapdf15}
\end{center}
\end{figure}

These analyses still have some dependence on the PDF used to predict the jet cross sections. 
This can be better accounted for by making a simultaneous fit of PDFs and $
\alpha_s(M_Z)$. This was done by the HERA experiments by using H1 and ZEUS
 jet data together with the combined HERA inclusive data.
The preliminary HERA-II data were first combined with 
the HERA-I data to yield an inclusive 
data set wih improved accuracy at high $Q^2$ and high $x$~\cite{highq2}. 
This new data set is used as the 
sole input to  a PDF fit called HERAPDF1.5~\cite{herapdf15} 
which uses the same formalism 
and assumptions as the HERAPDF1.0 fit. 
Fig.~\ref{fig:herapdf15} (left) shows the combined data for $NC$ $e^{\pm}p$ 
cross-sections with the HERAPDF1.5 fit superimposed. The parton distribution
functions from HERAPDF1.0 and HERAPDF1.5 are compared in 
Fig.~\ref{fig:herapdf15} (right). The improvement in precision at high $x$ 
is clearly visible.

The HERAPDF1.5 analysis has been extended to include H1 and ZEUS
 inclusive jet data~\cite{h1hq2,h1lq2,zeus97,zeus00}.
 The new PDF set which 
results is called HERAPDF1.6~\cite{herapdf16}.
For these fits the HERAPDF1.5 parametrisation of the gluon distribution and 
the valence distribution is extended. 
This more flexible parametrisation is called HERAPDF1.5f~\cite{jphysgrev}. 
The extra 
flexibility does not change the NLO PDFs central values significantly, and 
the PDF uncertainties are also not much increased.

\begin{figure}[htb]
\begin{tabular}{cc}
\includegraphics[width=0.45\textwidth]{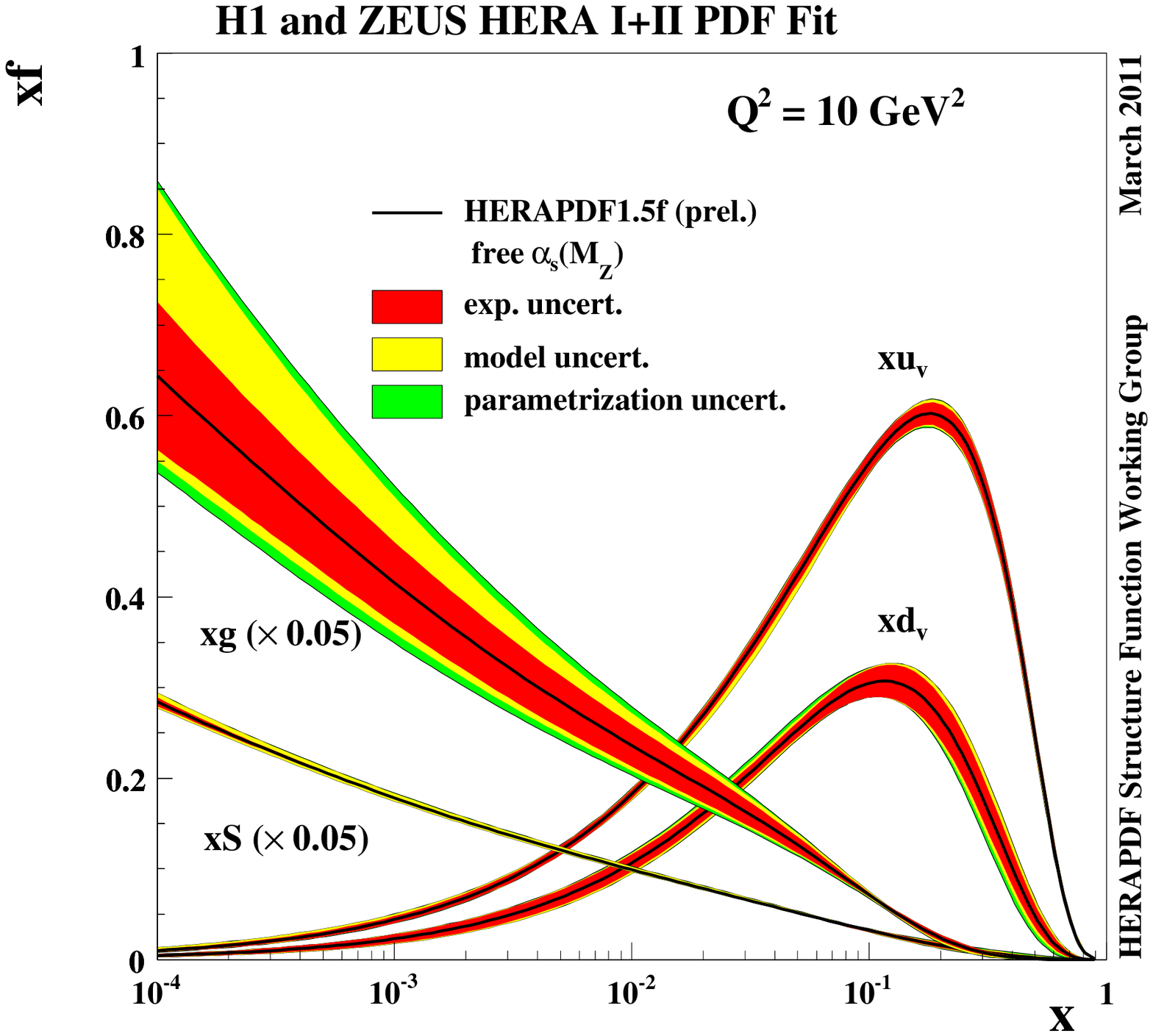} &
\includegraphics[width=0.45\textwidth]{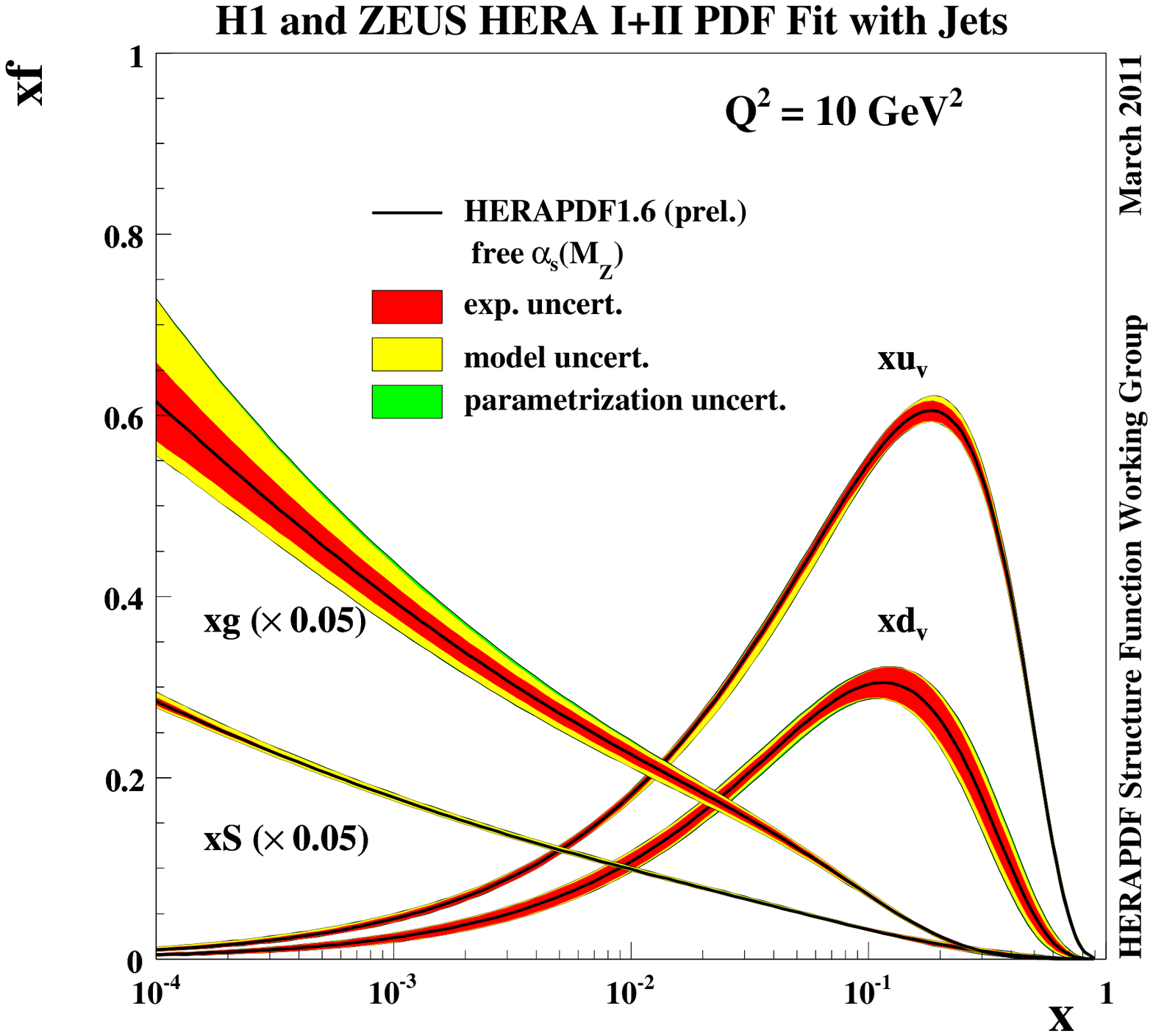}
\end{tabular}
\caption {The parton distribution functions
 $xu_v,xd_v,xS=2x(\bar{U}+\bar{D}),xg$, at $Q^2 = 10$~GeV$^2$, from 
HERAPDF1.5f and HERAPDf1.6, both with $\alpha_S(M_Z)$ 
treated as a free parameter of the fit.
The experimental, model and parametrisation 
uncertainties are shown separately. The gluon and sea 
distributions are scaled down by a factor $20$.
}
\label{fig:jetnojetalph}
\end{figure}

When the jet data are added the HERAPDF1.6 PDFs are 
also similar, and 
there is no tension between the jet data and the inclusive data.
For the jet data the NLO cross-sections are caluclatedusing 
NLOJet+~\cite{NLOjet+}.
 The impact of the jet data is clearly seen when 
$\alpha_S(M_Z)$ is allowed to be a free parameter of the fit. 
Fig.~\ref{fig:jetnojetalph} shows the PDFs
for HERAPDF1.5f and HERAPDF1.6, each with $\alpha_S(M_Z)$ left 
free in the fit. It can be seen that without jet data the uncertainty on 
the gluon PDF at low $x$ is large. This is because there is a strong 
correlation between the low-$x$ shape of
the gluon PDF and $\alpha_S(M_Z)$.  However once jet data are 
included the extra information on gluon induced processes reduces this 
correlation and the resulting 
uncertainty on the gluon PDF is not much larger than it 
is for fits with $\alpha_S(M_Z)$ fixed.

\begin{figure}[htb]
\begin{center} 
\includegraphics[width=0.5\textwidth]{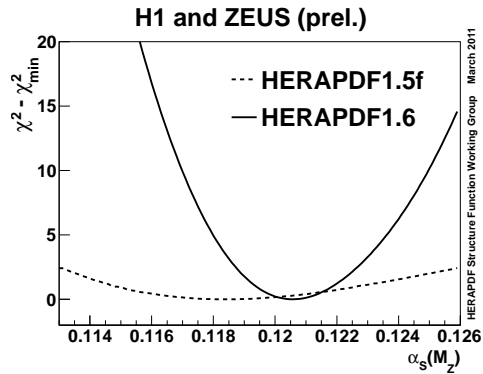} 
\caption {The difference between $\chi^2$ and its minimum value for the 
HERAPDF1.5f and HERAPDf1.6 fits as a function of $\alpha_s(M_Z)$
}
\end{center}
\label{fig:chiscan}
\end{figure}

Fig.~\ref{fig:chiscan}  shows a $\chi^2$ scan vs $\alpha_S(M_Z)$ for the fits 
with and without jets, illustrating how much better $\alpha_S(M_Z)$ is 
determined when jet data are included. The model and parametrisation errors 
are also much better controlled.

The value of $\alpha_s(M_Z)$ extracted from the HERAPDF1.6 fit is:

$
\alpha_S(M_Z) = 0.1202 \pm 0.0013(exp) \pm 0.0007(model/param) \pm 0.0012 (had) +0.0045/-0.0036(scale)
$

Model and parametrisation uncertainties on $\alpha_S(M_Z)$ are estimated
in the same way 
as for the PDFs~\cite{jphysgrev} and the uncertainties on the hadronisation 
corrections applied to the jets are also evaluated. 
The scale uncertainties are estimated by 
varying the renormalisation and factorisation scales chosen in the jet 
publications by a factor of two up and down. The dominant contribution to the 
uncertainty comes 
from the jet renormalisation scale variation.

\section{Summary}
Charm production data from the HERA experiments ZEUS and H1 have been 
combined and used to determine the charm quark mass within various heavy 
quark schemes. Jet cross sections at HERA have been used to determine $\asmz$.

\end{document}